\documentclass{iaus}
\usepackage{graphicx}

\title[The ALFALFA Search for (Almost) Dark Galaxies] 
{The ALFALFA Search for (Almost) Dark Galaxies across the HI Mass Function}

\author[M. P. Haynes]   
{Martha P. Haynes$^1$}

\affiliation{$^1$ Center for Radiophysics and Space Research,
Cornell University, Ithaca, NY 14853, USA \break email: haynes@astro.cornell.edu}

\pubyear{2007}
\volume{244}  
\pagerange{xxx-xxx}
\date{?? and in revised form ??}
\jname{Dark Galaxies and Lost Baryons}
\editors{J.I.Davies \& M.J. Disney, eds.}

\def\kms{km~s$^{-1}$}
\def\etal{{\it et al.}}
\def\sqd{{deg$^{2}$}}
\def\arcmin{$^{\prime}$}
\def\arcsec{$^{\prime\prime}$}
\def\msun{$M_\odot$}
\def\mhi{$M_{HI}$}
\def\lsun{$L_\odot$}

\begin{document}

\maketitle

\begin{abstract}
The Arecibo Legacy Fast ALFA (ALFALFA) survey is a second
generation blind extragalactic HI survey currently in progess
which is exploiting Arecibo's superior sensitivity, angular resolution
and digital technology to derive a 
census of the local HI universe over a cosmologically significant volume.
As of the time of this meeting, some 4500 good quality extragalactic
HI line sources have been identified in about 15\% of the final survey area.
ALFALFA is detecting HI masses as low as 10$^6$\msun ~and
as large as 10$^{10.8}$\msun ~with positional accuracies typically better 
than 20\arcsec, allowing immediate identification of the most probable
optical counterparts. Only 3\% of all extragalactic HI sources
and less than 1\% of detections with 
\mhi ~$> 10^{9.5}$\msun ~cannot be identified with a stellar component.
Because ALFALFA is far from complete, the discussion
here focuses on limitations of past surveys that ALFALFA will overcome
because of its greater volume, sensitivity and reduced susceptibility
to source confusion and on a sampling of illustrative preliminary results.
First ALFALFA results already suggest, in agreement with previous studies, 
that there does not appear to be a cosmologically significant population of 
optically dark but HI rich galaxies.  ALFALFA promises
a wealthy dataset for the exploration of many issues in near-field cosmology 
and galaxy evolution studies, setting the stage for their extension
to higher redshifts in the future with the Square Kilometer Array (SKA).
\end{abstract}

\firstsection 
\section{Introduction}

As described in the papers by Giovanelli, Martin and others in these
proceedings, the Arecibo Legacy Fast ALFA (ALFALFA) survey is exploiting the
superior sensitivity, angular resolution, spectrometer and signal
processing associated with the Arecibo L--band Feed Array (ALFA) 
on the Arecibo 305~m antenna to conduct a ``second generation'' wide 
area blind HI survey. Following on the success of the
 HI Parkes All-Sky Survey (HIPASS:
Zwaan \etal ~2004; Meyer \etal ~2004), ALFALFA offers substantial gains
with respect to HIPASS (as it should!), most notably in its increased
depth and angular resolution. With a median redshift of only $\sim$2800 \kms,
HIPASS did not sample adequate extragalactic volume to yield a 
cosmologically ``fair sample'' of the local universe, and the large beamsize
(15\arcmin) of the Parkes telescope made identification of optical 
counterparts often uncertain without followup HI synthesis observations. In fact,
Oosterloo \etal ~(2007) suggest that confusion within the large Parkes
beam creates significant uncertainty in the estimate of statistical
properties based on HIPASS. With 20 times smaller beam area and eight
times the sensitivity, the centroiding accuracy of ALFALFA is
on average 24\arcsec ~(20\arcsec ~median) for all sources with
signal-to-noise ratio $> 6.5$, and its median redshift is nearly
three times greater (Giovanelli \etal ~2007). Thus, ALFALFA 
will produce a deeper and richer census of the local HI universe, 
probing a cosmologically significant volume and, in most instances,
providing immediate identification of the corresponding optical
counterpart to each HI source. 
Because ALFALFA is in progress,
its results are not fully available. However, 
the currently available ALFALFA catalog contains $\sim$4500
good quality extragalactic HI detections so that suggestive preliminary
results are beginning to emerge. In the context of the present
discussion of ``Dark Galaxies and Lost Baryons'', the most interesting
objects will be ones without a clearly identified stellar component.
Of the current sample, only 3\% cannot be identified with a distinct 
optical counterpart and virtually all of the objects with \mhi ~$> 10^{9.5}$ \msun
~can be associated with a luminous galaxy in standard imaging
datasets such as SDSS or 2MASS. The majority of low mass HI detections 
without discrete stellar counterparts can be explained as debris
remnants of a past interaction involving a nearby optically-seen
galaxy (Kent \etal ~2007; Haynes, Giovanelli \& Kent 2007).
In this paper, I summarize the potential for ALFALFA to deliver 
cosmologically significant results and present some illustrative findings
which the full survey promises to address in more quantitative detail.

\section{Determinations of the HI Mass Function}\label{sec:HIMF}

One of the principal discrepancies between cold dark matter (CDM) theory and
current observations revolves around the difference between the expected and observed
numbers of low mass dark matter halos (Kauffmann, White \& Guiderdoni ~1993; Klypin \etal ~1999), 
sometimes referred to as the ``missing satellite problem''. 
The logarithmic slope of the faint end of the mass function predicted by CDM
simulations is close to the Press-Schechter (Press \& Schechter 1974) value 
of $\alpha = -1.8$. Because the mass function itself
is difficult to determine directly, observational efforts have mainly focused
on estimates of the faint end of the optical luminosity function and 
its HI counterpart, the HI mass function (HIMF). By determining
both, limits can be set on the number of low mass halos containing measurable
stellar or gaseous components. The shape of the low mass end of the HIMF
and its corollary, the cosmological mass density of HI, are important
parameters in the modeling of the formation and evolution of galaxies.

The first determinations of the HIMF (e.g.,
Zwaan \etal ~1997) used the venerable $\Sigma(1/V_{max}$)
method (Schmidt 1964).  Application of that method 
assumes that the HI sources are
homogeneously distributed within the sample volume, an assumption
that is obviously incorrect. An alternative method for
deriving luminosity and mass functions, the two-dimensional
stepwise maximum likelihood (2DSWML) method proposed by
Loveday (2000) has been employed by Zwaan \etal ~(2003) and
Springob \etal ~(2005a). The 2DSWML method also has drawbacks
in that it is not automatically normalized and assumes that the
shape of the HIMF is universal. In practice, both methods are obviously
limited by statistics; they can also be severely affected by 
systematic uncertainties, such as errors in the estimate
of distances, especially nearby (Masters \etal ~2004).
Adopting a modified $\Sigma(1/V_{max}$) approach, Rosenberg
\& Schneider (2002) and Springob \etal ~(2005a) accounted in different
ways for the effects of large scale structure
by assigning an effective weight for each galaxy to account
for density inhomogeneities.
As the latter authors point out, a survey which does not
sample adequate volume may produce unreliable results because
the weighting scheme must account not only for variations
of the space density of galaxies with distance but also
for variations in the fraction of the surveyed volume
that is occupied by regions of a particular density.

While blind HI surveys avoid issues associated with selection
bias favoring high optical surface brightness targets, the fact that
there does not appear to be a significant population of optically
dark galaxies at least at the higher masses suggests that a significantly
large and deep optically-selected sample could detect enough HI galaxies
to uncover the true HIMF of spiral galaxies at least in the high mass
regime. Springob \etal ~(2005a) have derived the HIMF and examined its 
dependence on morphological type and local environment
for a catalog of $\sim$8800 galaxies detected
in HI through optically-targeted observations (Springob \etal ~2005b).
Because of the size and depth of their sample, Springob \etal ~were able to 
identify a strictly diameter- and flux-limited subset, still with $\sim$2800
objects in it. They used both a modified $\Sigma(1/V_{max}$) method,
adopting a weighting scheme based on the $PSCz$ density field
(Branchini \etal ~1999), and the 2DSWML method, achieving results
that agree within the derived uncertainties.

\begin{figure}[t]
\centering
\resizebox{6.0cm}{!}{\includegraphics{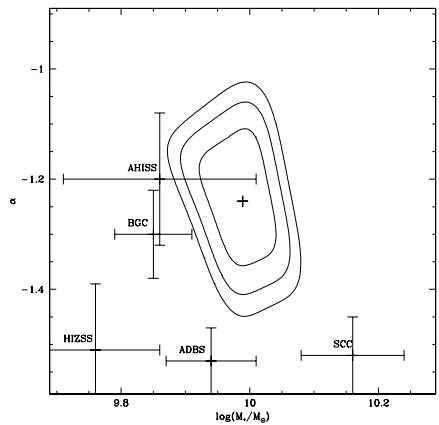}}
\vskip -1.2cm
\caption{$\chi^2$ contours of the Schechter fit parameters
$M_*$ and $\alpha$ to the HIMF for the Springob \etal ~(2005a)
optically-selected sample. Values with error bars indicate fits
derived from the HI blind surveys: HIZSS (Henning \etal ~2000),
BGC (Zwaan \etal ~2003), AHISS (Zwaan \etal ~1997),
ADBS (Rosenberg \& Schneider 2002) and SCC (Kilborn 2000). See
Springob \etal ~(2005a) for further details.}
\label{fig:schechter}
\end{figure}

To date, determinations of the HIMF by both blind and targeted
HI surveys have produced similar but not entirely reconcilable
results. Figure \ref{fig:schechter} shows a comparison of the derived
best-fit Schechter function characteristic HI mass $M_*$ and
low mass end slope $\alpha$ for a selection of recent determinations
of the HI mass function. The contours outline the 1-, 2- and 3-$\sigma$
fits to the Springob \etal ~(2005a) HIMF. The agreement between
the parameter fits derived from that optically selected sample 
and those derived from the HI blind surveys is no worse than
the internal agreement among the latter. Evident also in early results 
derived from limited optical redshift surveys of the local universe, 
discrepancies among the HIMFs derived from current HI surveys are 
clear symptoms of the lack of a fair sample. While optically
selected samples may exclude gas rich, low optical luminosity 
and low surface brightness galaxies, the HI blind surveys preceding
ALFALFA have sampled a much shallower volume than typical optical surveys.
The sampling of insufficient depth raises the potential for 
systematic errors in the HI mass estimates due to distance uncertainties
and in the selection of a population of galaxies that is not representative
of the global population detected at larger distances. For example,
Masters \etal ~(2004) have shown the the slope of the HIPASS
HIMF is actually steeper than reported by
Zwaan \etal ~(2003) when the effects of peculiar velocities in the
surveyed volume are properly taken into account. Furthermore,
the lack of sufficient sampling of galaxies with very high HI masses,
\mhi ~$> 10^{10}$\msun, reduces the predictive power of past
surveys to expectations for future surveys of HI emission from galaxies
at moderate to high redshift, as discussed in Section \ref{sec:highmass}.

\begin{figure}[t]
\centering
\resizebox{10.0cm}{!}{\includegraphics{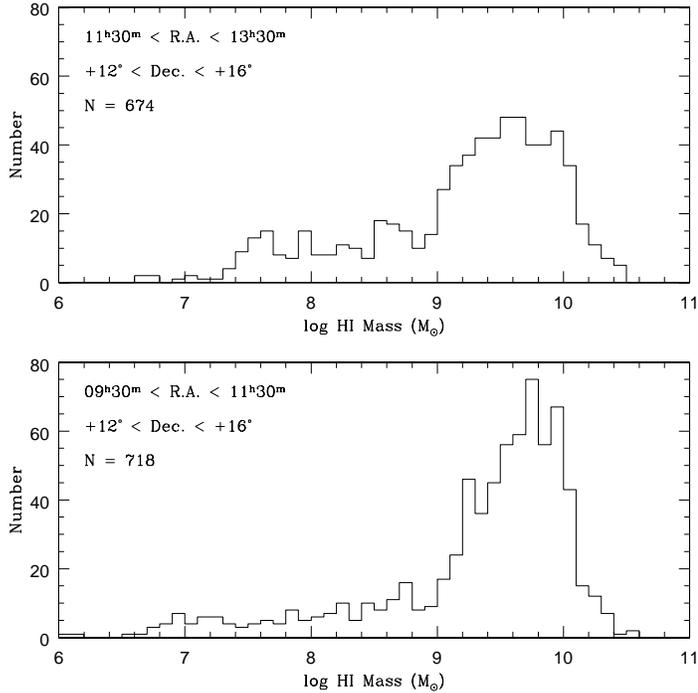}}
\caption{Left: Upper: Histogram of the HI masses for ALFALFA sources
in the region $11^h30^m <$ R.A. $< 13^h30^m$, $+12^\circ <$ Dec. 
$< +16^\circ$ including the Virgo cluster region.
All objects with c$z < 3000$ \kms ~and found within 5$^\circ$
of M87 are placed at the Virgo distance
of 16.7 Mpc.. Lower: Similar histogram 
over the same declination band but in the range 
$09^h30^m <$ R.A. $< 11^h30^m$.
}\label{fig:histo}
\end{figure}

\section{The Low Mass End of the HIMF}\label{sec:lowmass}

Previous determinations of the HIMF below 10$^8$ \msun ~(Zwaan \etal
~1997; Rosenberg \& Schneider 2002; Zwaan \etal ~2003) suffer severely
from small number statistics and from the systematics associated with
distance uncertainties and large scale inhomogeneities. 
As discussed in Section 4 of Giovanelli \etal ~(2005a), the most
advantageous approach to increasing the number of detections of
a given HI mass (once such a mass is
detectable at an astrophysically interesting
distance) in a survey of fixed observing time
is to increase the solid angle of the survey, not its
depth. Hence the Arecibo Galaxy Environments Survey (AGES) discussed
by Minchin in this volume, will have a median redshift
1.5 times that of ALFALFA and will explore richly selected
regions of nearby clusters and groups, but AGES will detect far 
fewer low mass objects than ALFALFA will because it covers only 
1/35th as much solid angle. 

ALFALFA is specifically designed to survey enough solid angle to
detect several hundred objects with \mhi ~$< 10^{7.5}$ and thus provide
a robust determination of the low HI mass slope of the HIMF. 
Already, ALFALFA has detected more galaxies with masses less than
\mhi ~$< 10^{7.5}$ \msun ~than included in all of the previous
HI blind surveys combined.

A major issue confronting the determination of the low mass end of
the HIMF arises from the uncertainty in distance estimates derived
from redshift information, even when sophisticated flow models based
on redshift-independent distances are used (Masters \etal ~2004).
Using the current ALFALFA dataset,
Figure \ref{fig:histo} illustrates the resultant problem. 
The upper panel shows the histogram of HI masses
contained in a 120 \sqd ~region from  $11^h30^m <$ R.A. $< 13^h30^m$, 
$+12^\circ <$ Dec. $< +16^\circ$. There is no restriction placed
on redshift, but galaxies found within 5$^\circ$ of M87 and
with c$z < 3000$ \kms ~are assumed to be members of the
Virgo cluster and are assigned distances of 16.7 Mpc. Elsewhere,
a flow model derived from a combination of primary and secondary
distances is used (Masters 2005). Very few galaxies are detected
with \mhi ~$< 10^{7.5}$ \msun. For comparison, the lower panel
shows a similar histogram for a contiguous equal area
extending from  $09^h30^m <$ R.A. $< 11^h30^m$ at the same declinations.
The difference in the histograms is striking. At higher masses, the
differences can be ascribed to the known HI deficiency of galaxies
in Virgo and cosmic variance in the large scale structure sampled 
by the two regions. However, it is likely that
the apparent deficit of galaxies at lowest HI mass and the
bulge at \mhi ~$\sim 10^{7.9}$ \msun ~seen in the upper panel
can be explained in part by the incorrect assignment of distances
to some of the galaxies in the vicinity of Virgo.
If some of them actually lie in the foreground,
then their masses are overestimated; use of the correct, nearer
distances would shift some of the objects in the peak
to fill in those ``missing'' at lower masses. Future work
to obtain more precise distance estimates for these objects is
clearly in order both to explore physical mechanisms which may
alter the HIMF locally and to understand
better the large scale structure in and around the Virgo cluster.

\begin{figure}[t]
\centering
\resizebox{13.0cm}{!}{\includegraphics{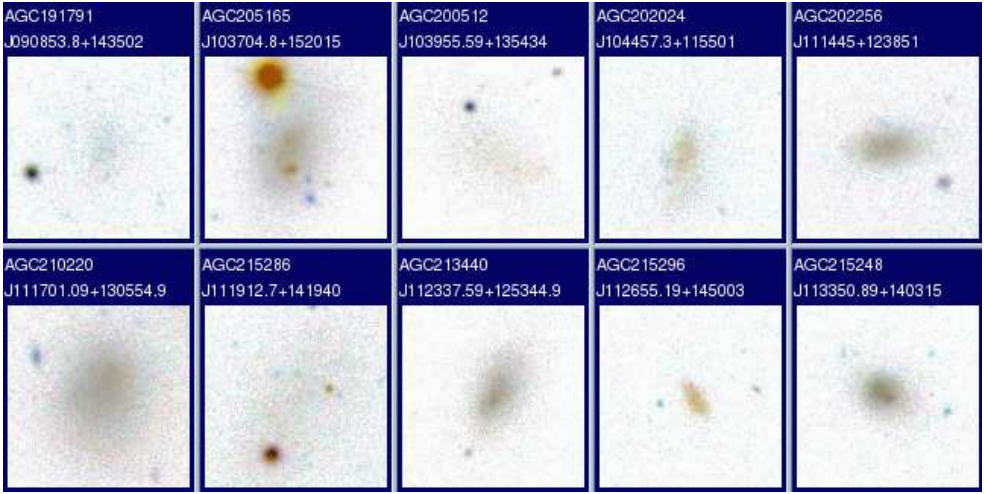}}
\caption{SDSS images of the optical counterparts of representative
lowest HI mass (\mhi ~$< 10^{7.2}$ \msun)
ALFALFA detections. Each image is 50\arcsec ~square and uses
standard SDSS Sky Server scaling. }\label{fig:lowmass}
\end{figure}

\subsection{The Low Mass Galaxies as (Almost) Dark Galaxies}
\label{sec:dwarfs}

In addition to providing a robust determination of the faint
end slope of the HIMF, ALFALFA promises to identify a sample
of nearby, low mass dwarf galaxies which can be identified
from more massive but gas--poor systems because of their
small HI line widths. Figure \ref{fig:lowmass} shows the SDSS
images of a representative set of the lowest HI mass systems,
with \mhi ~$< 10^{7.2}$ \msun. All of them are nearby,
optically faint and of low surface brightness. They exhibit
a range of morphological clumpiness and evidence of massive
star formation.

\begin{figure}[t]
\centering
\resizebox{14.0cm}{!}{\includegraphics{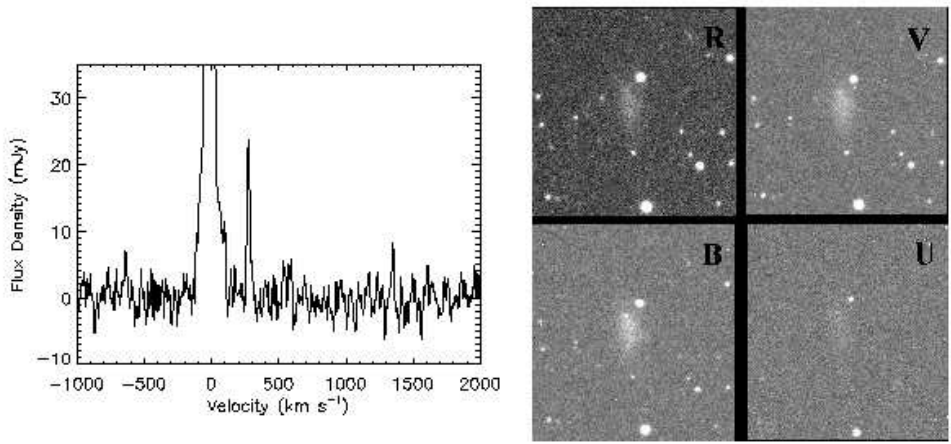}}
\caption{Left: ALFALFA spectrum of AGC~112521 (HI0141+27). Right: 
Broad-band images (courtesy of L. van Zee) using the WIYN 0.9m 
telescope of the optical counterpart, a newly-discovered 
low surface brightness dwarf member of the NGC~672 group.}\label{fig:HI0141}
\end{figure}

For her Ph.D. thesis, Am\'elie Saintonge 
(2007a,b) has undertaken a first study of some of the lowest 
HI mass galaxies detected early in the ALFALFA survey. Various
members of the ALFALFA consortium are pursuing multiwavelength
observations including broad-band and H$\alpha$ imaging, 
HI synthesis mapping and UV imaging with the GALEX satellite.
Of a first sample targeted for follow-up H$\alpha$ imaging,
15\% do not show the presence of HII regions.
While this work is still on-going, we present an illustrative
example of the kind of newly catalogued system that ALFALFA
finds. Figure \ref{fig:HI0141} shows the ALFALFA HI spectrum of
a new dwarf member of the NGC~672 group dubbed AGC~112521.
This object was first
detected in our ALFA-commissioning
precursor observations (Giovanelli \etal ~2005b) and recovered
in the early ALFALFA survey observations. As evident in Figure 
\ref{fig:HI0141}, the HI line emission from the galaxy is clearly detected
at c$z = +274$ \kms, a narrow line full width at 50\% of the peak 
emission of 26 \kms and a total HI line flux density of 0.68
Jy-\kms ~(Saintonge \etal ~2007). Assuming its membership in the NGC~672
group at a distance of 7.2 Mpc (Sohn \& Davidge 1996; Karachentsev \etal ~2004),
the HI mass is $7.9 \times 10^6$ \msun. Its blue luminosity L$_B$ derived
from newly acquired images is similarly low L$_B = 3.6  \times 10^6$ \lsun,
so that \mhi/L$_B$ is 2.18. Adopting standard relations to convert multiband
magnitudes into stellar mass, we find that nearly half of the baryonic matter
is still in the form of gas.

With the promise of a final catalog 
of thousands of galaxies in the mass range $10^6 <$ \mhi ~$< 10^8$ \msun,
the study of these gas rich systems is a prime science driver of ALFALFA.
Future ALFALFA studies will explore not only the low mass slope of the HI
mass function and its possible dependence on environment, but also
the distribution, morphology, star formation rate and chemical
enrichment history of these low mass gas-rich dwarfs.

\section{The Environment and Clustering of Gas-Rich Galaxies}
\label{sec:cluster}

In addition to the missing satellite problem, the behavior of the
HIMF with varying galaxy environment has a direct bearing on 
the use of the HI line for cosmological purposes and on
fundamental issues of galaxy formation and evolution.
Early studies of the possible
environmental dependence of the HIMF were limited to comparisons
of the HIMF derived for galaxies in the Virgo cluster with that derived
for galaxies in the
field (Hoffman \etal ~1992; Briggs \& Rao 1993; Rosenberg \& Schneider
2002; Davies \etal ~2004; Gavazzi \etal ~2004). While all of these studies
suffer to varying degrees from poor statistics and incompleteness,
their results marginally suggest that the HIMF in Virgo is missing the low
HI mass dwarfs found in the field or is at least flatter at the faint end
than the field HIMF.  Exploiting the large and deep optically-selected sample
of Springob \etal ~(2005b), Springob \etal ~(2005a), divided the sampled
volume into density regimes based on the $PSCz$ (Branchini 
\etal ~1999) density field and compared the derived HIMFs in different
regimes. They concluded
that the low mass slope $\alpha$ is shallower and $M_*$ lower
in high density environments, in qualitative agreement with the suggestions
of the earlier studies based on Virgo alone.
Zwaan \etal ~(2005) also looked at the environmental dependence of
the HIMF and came to the opposite conclusion: that the HIMF
is steeper in high density regions. However, those authors use
a local density estimator that is purely HIPASS-based, that is,
reflective of the density field of the survey itself, with no
corrections for peculiar velocities or bias. Other recent work,
based on the Canes Venatici region, confirms the earlier
findings that the low mass slope is shallow in that region
(Kova\u{c}, Oosterloo \& van der Hulst 2005). 
A much larger sample with many more galaxies over a larger
volume is required in order to allow an investigation
of whether the HIMF shape is dependent on morphological type and 
environment separately, as has been done by Croton 
\etal ~(2005) for the 2dF Survey optical luminosity function.
Clearly ALFALFA will contribute 
significantly to a firmer understanding of the interplay between 
environment and the HIMF.

Although they acknowledge the volume and resolution limitations of the HIPASS catalog, 
recent papers have explored the behavior of the galaxy distribution it traces by
attempting to estimate its two--point correlation function $\xi (r)$. Meyer
\etal ~(2007) conclude that gas-rich galaxies are among the most weakly
clustered galaxies known and suggest that the clustering scale length
$r_o$ depends strongly on rotational velocity, and thus, by implication,
on the halo mass. In contrast, Basilakos \etal ~(2007)
argue that massive HIPASS galaxies show the same clustering properties
as optically-selected ones, but that
the low mass systems, \mhi $< 10^9$ \msun, show a nearly uniform distribution.
Both of these studies are fraught with potential systematics because
of the shallow depth of HIPASS, the potential impact of source confusion
and the statistical limitations when 
its catalog is divided into subsets. As the follow-on to HIPASS,
ALFALFA is specifically designed to overcome many of the limitations of 
the earlier survey, thereby allowing a robust determination of the
HI-HI and HI-optical galaxy correlation functions and a quantitative study
of the biasing of the HI population relative to optical or IR-selected
samples and to the underlying density field. 

\subsection{The Void Problem}\label{sec:voids}

As pointed out by Peebles (2001), numerical simulations based
on CDM predict that voids should contain large numbers of dwarf galaxies.
Peebles suggests that the failure to identify such a void population 
raises one of the principal challenges to CDM models.
For example, the voids in the simulations of 
Gottl\"ober \etal ~(2003) are criss-crossed by dark matter filaments,
within which lie large numbers of very low amplitude inhomogeneities.
In fact, those authors predict that a void with a diameter of 20$h^{-1}$
Mpc should contain 1000 dark matter halos with masses of
10$^9$ \msun ~and as many as 50 with masses ten times greater than that.
Photoionization and baryonic blowout may suppress star formation or
perhaps the retention of any baryons within the low mass halos, but 
it is not at all clear
that these processes are sufficient to explain the absence of galaxies 
in voids (Hoeft \etal  ~2006).

Limits on the abundance and properties of void galaxies have been
placed by previous optical and radio studies.
Hoyle \etal ~(2005) have used the SDSS dataset to show that
the luminosity function of void galaxies has a fainter break 
luminosity L$^*$ but a similar faint end slope to the overall 
SDSS luminosity function. In addition, they have shown that
void galaxies are typically blue, disk-like and have
high H$\alpha$ equivalent widths, making them excellent targets
for HI emission line surveys.

Previous surveys for HI in voids have exploited the VLA to conduct
blind HI surveys of the Pisces-Supercluster and its foreground void 
(Weinberg \etal ~1991; Szomoru \etal ~1994) and the
Bootes void (Szomoru \etal ~1993; 1996). 
Szomoru \etal ~(1993) did find an isolated galaxy in Bootes, but its
mass of $5 \times 10^9$ \msun ~and blue luminosity
L$_B$ of $-18.6$ exclude it as a true dwarf. In fact, those VLA surveys
sampled a relatively small volume and were hampered in the detection
of low mass objects by poor spectral resolution (42 \kms). More
recently, Pustilnik \etal ~(2002) have explored the HI content
of a sample of blue compact galaxies known to be located in voids
and find them to be ``darker'', as measured by their HI mass 
to blue luminosity ratios \mhi/$L_B$  than their counterparts in higher
density regions.

Saintonge \etal ~(2007) initiated a first, but limited, ALFALFA
analysis of galaxies in the
nearby void in front of the Pisces-Perseus foreground void 
at $cz \sim 2000$ \kms. They detected no galaxies
in a large volume of 460 Mpc$^3$, whereas a scaling of the 
predictions of Gottl\"ober \etal ~(2003) under the assumption 
that the dark-to-HI mass ratio is 10:1 predicts that ALFALFA
would have detected 38 HI sources. This very preliminary
result for a single void in only 2\% of the ALFALFA survey suggests that
the discrepancy between the predicted and observed abundance of
dwarf galaxies in voids cannot be reconciled by a population of
gas rich dwarfs. More credible results will be available when the
full ALFALFA
survey is completed.

\section{The High Mass End: Prelude to the SKA}\label{sec:highmass}

\begin{figure}[t]
\centering
\resizebox{13.0cm}{!}{\includegraphics{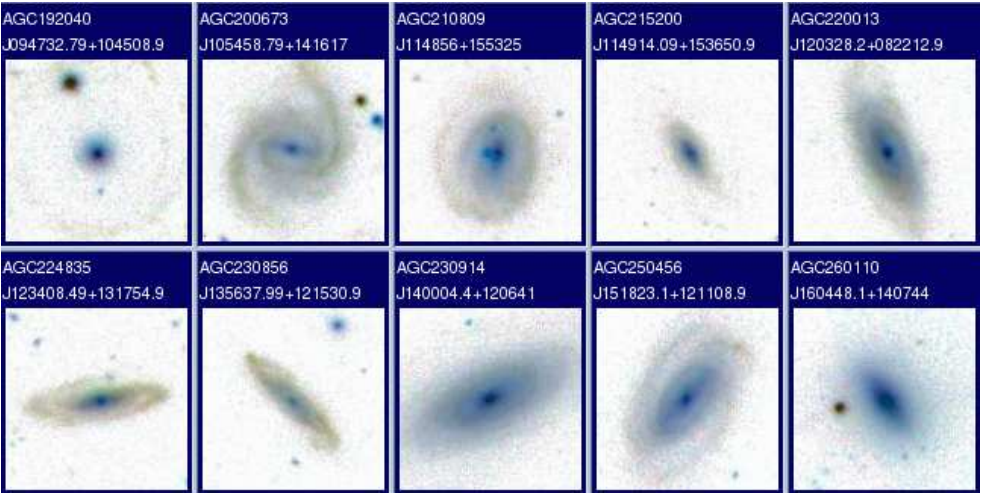}}
\caption{SDSS images of the optical counterparts of representative
highest HI mass (\mhi ~$> 10^{10.4}$ \msun)
ALFALFA detections. Each image is 50\arcsec ~square and uses
standard SDSS Sky Server scaling. Massive HI galaxies such as these
will be detected in large numbers by planned future SKA surveys.}
\label{fig:highmass}
\end{figure}

One of the prime science drivers of the Square Kilometer Array (SKA)
is the undertaking of a billion galaxy redshift survey in the HI line
over the redshift range $0 < z < 2.5$ to explore the evolution
of the gas content of galaxies and constrain the dark energy equation
of state through the measurement of baryon acoustic oscillations
(Abdalla \& Rawlings 2004). Even allowing for the likely
increase in the gas content with $z$, only the most massive HI galaxies 
will be detected in emission at moderate redshift. And, as in the 
case of the low mass end of the HIMF, previous surveys have been
too shallow to detect the most massive HI galaxies. ALFALFA
has already detected more than twice as many massive galaxies with
\mhi ~$> 10^{10.4}$ \msun ~than all the previous HI blind surveys
combined. Figure \ref{fig:highmass} shows SDSS images of the
optical counterparts of a representative set of these massive ALFALFA
detections. They exhibit a range of morphologies, colors and nuclear
concentration but all appear to be luminous disk systems. Many of these
have stellar masses in the range corresponding to the ``transition mass''
($M_{stars} \sim 3 \times 10^{10}$ \msun) above which galaxies 
show a marked decrease
in their present to past-averaged star formation rates (Kauffmann \etal
~2003). 

ALFALFA will
contribute its high mass detections to the GALEX--Arecibo--SDSS Survey 
(GASS; P.I.: D. Schiminovich) which aims to obtain measures
of the HI content at a fixed gas mass fraction (1.5\%) of the
stellar mass in a sample of 1000 galaxies chosen by optical-UV
criteria to have stellar masses $> 10^{10}$ \msun. 
The combination of multiwavelength data will
provide new understanding of the physical processes that regulate
gas accretion and its conversion into stars in massive systems.
HI can currently be detected in emission from normal galaxies
out to a redshift approaching $z \sim 0.3$. The highest redshift
detection of a single object to date, reported by Catinella
in this volume, is $z \sim 0.28$; such observations require several
hours of telescope time with Arecibo today. While ALFALFA$+$GASS will
characterize the properties of galaxies at $z \sim 0$, ambitious
future studies aimed at characterizing the evolution of galaxies
over the last 4 Gyr should be possible in the next few years
with Arecibo as well as the SKA precursor instruments under
construction today.

\section{Conclusions}\label{sec:concl}

ALFALFA is an ongoing survey with a detection catalog available to us
at the time of this conference amounting to only about 15\% of the final survey.
Given its state, the full impact of ALFALFA is only beginning to
become evident, but the survey promises now to yield $> 25000$ extragalactic
HI detections when it is complete. It should yield robust measures of the
HIMF, the HI-HI and HI-optical correlation functions and their bias
parameters at $z = 0$, thereby laying a firm footing for future studies
of their evolution over cosmic time.
ALFALFA is an open consortium and interested
parties are invited to follow the survey's progress via the ALFALFA website
{\it http://egg.astro.cornell.edu/alfalfa}.

\begin{acknowledgments}
This work has been supported by NSF grants AST--0307661,
AST--0435697 and AST--0607007 and by the Brinson Foundation. 
The Arecibo Observatory is part of the National Astronomy and Ionosphere
Center which is operated by Cornell University under a cooperative
agreement with the National Science Foundation.
Funding for the SDSS and SDSS-II has been provided by the Alfred P. 
Sloan Foundation, the Participating Institutions, the National Science 
Foundation, the U.S. Department of Energy, the National Aeronautics 
and Space Administration, the Japanese Monbukagakusho, the Max Planck 
Society, and the Higher Education Funding Council for England. The 
SDSS Web Site is http://www.sdss.org/.

\end{acknowledgments}

\end{document}